# Yield stress and shear-banding in granular suspensions

Abdoulaye Fall[1,2], François Bertrand[2], Guillaume Ovarlez[2] and Daniel Bonn[1,3]
[1]*Van der Waals – Zeeman Institute, University of Amsterdam, Valckenierstraat 65, 1018XE Amsterdam, The Netherlands*
[2]*Laboratoire des Matériaux et Structure du Génie Civil, 2 Allée Kepler 77420 Champs sur Marne, France*
[3]*Laboratoire de Physique Statistique, Ecole Normale Supérieure, 24 Rue Lhomond 75231Paris Cedex 05, France*

We study the emergence of a yield stress in dense suspensions of non-Brownian particles, by combining local velocity and concentration measurements using Magnetic Resonance Imaging with macroscopic rheometric experiments. We show that the competition between gravity and viscous stresses is at the origin of the development of a yield stress in these systems at relatively low volume fractions. Moreover, it is accompanied by a shear banding phenomenon that is the signature of this competition. However, if the system is carefully density matched, no yield stress is encountered until a volume fraction of 62.7 ± 0.3%.

According to one of the standard textbooks on granular materials [1], the processing of granular materials consumes roughly 10% of all the energy consumed on this planet. Consequently the prediction of flow resistance of granular materials is a matter of considerable importance. The two simplest cases of granular materials are dry and wet sand. The former has recently been studied in much detail, and its flow behavior is by now well understood [2].

The other case, suspensions of (noncolloidal) granular particles in Newtonian fluids ("granular suspensions") should *a priori* be simple systems as the only interactions between the particles are hydrodynamic and contact forces. However, they exhibit a very rich behavior: yield stress [2-5], shear banding [4-6], shear thickening [7], normal stresses [8,9] and shear-induced migration [10,5] that remain incompletely understood.

Perhaps the most important issue is the correct determination of the yield stress, a minimum stress to enforce a quastistatic flow. This is a matter of much current interest, as it has large repercussions on our understanding of complex fluid flows in general [11]. For vanishing flow speeds, hydrodynamic interactions are expected to play no role and the behavior of dry grains is recovered [3,4,12]. A frictional yield stress $\tau_c = \mu\sigma_N$ is then observed provided the granular skeleton, of macroscopic friction coefficient $\mu$, is subject to a normal stress $\sigma_N$ [3,4,12]. However, this predicts the absence of a yield stress without normal forces, which is not what was observed experimentally e.g. in [4].

Another important question is the origin of yield stress observed in rather loose suspensions, for low volume fractions ranging from 55 to 60% [5]: what is the packing density at which a yield stress emerges? Theoretically this is expected to correspond to point J [13] i.e. to the density at which the granular skeleton becomes so densely packed that it can no longer flow. For frictionless granular materials this happens at a volume fraction of approximately $\varphi$=64%, corresponding to random close packing [14]. In suspensions, although some light has been shed on the role of friction [14] the role of the interstitial fluid and of contact lubrication remains an open question, and it is not clear at which volume fraction a yield stress develops in practice.

Equally puzzling is the observation of the coexistence between sheared and unsheared zones, characteristic of shear banding, during the viscous flows of such systems [4,5,15]. It was observed that there exist a critical shear rate below which no stable flows exist [4], which contradicts the commonly accepted wisdom that shear banding is due to heterogeneities in the stress field [11]. Shear-banding due to the existence of a critical shear rate has only been observed in thixotropic gels [8,11,16-18], and is not *a priori* expected in an athermal granular system with only hydrodynamic and contact forces.

In this Letter, we study the rheology of dense suspensions of non-Brownian particles and show that most if not all of the above problems disappear if one takes gravity into account. We find that it is due to gravity that a yield stress develops well below close packing. We also evidence that the critical shear rate for shear banding arises naturally as the results of the gravity and viscous stresses. And only if the particles and solvent are very carefully density matched, does a macroscopic rigidity develops very close to (but at a significantly smaller value than) the maximum random packing fraction of the granular matrix.

We study the local and global rheological behavior of a dense suspension of noncolloidal spherical particles immersed in a Newtonian fluid for 58%< $\varphi$ < 63%; most representative results presented here are obtained at 60%. We use polystyrene beads (diameter 40µm, polydispersity < 5%, density 1.05g.cm$^{-3}$). Suspensions are prepared by mixing the particles with aqueous solutions of NaI to perfectly match solvent and particle densities; this also allows to tune the density difference. For most experiments presented, the solvent is denser than the particles, and the particles cream rather than sediment.

Macroscopic rheometric experiments are performed with a vane-in-cup geometry (inner radius $R_i$ = 12.5mm, outer $R_e$ = 18.5mm) on a Bohlin 200 rheometer. MRI rheometry is performed in a wide-gap Couette geometry ($R_i$ = 41.5 mm, $R_e$ = 60mm). Local velocity and concentration profiles in the flowing sample were obtained through techniques described in detail in [19]; $\varphi$ can be obtained with an accuracy of 0.2% by measuring the local density of protons from the water [5]. The velocity profiles can also be measured [19,20]. For all experiments, in order to avoid slip at the walls, sandpaper is glued on the walls; on the velocity profiles there is no observable slip.

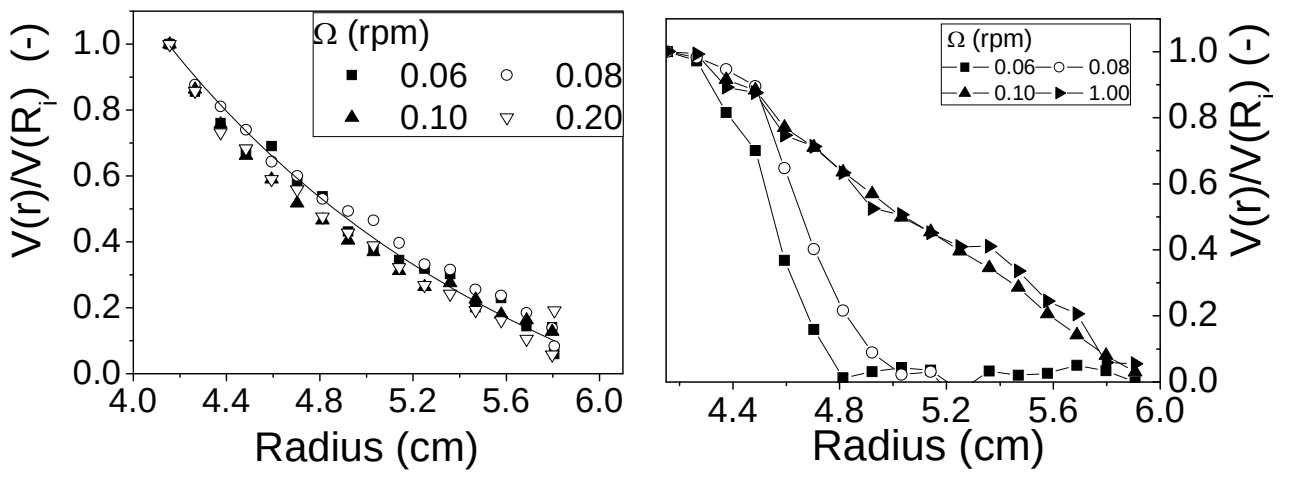

Fig. 1: Dimensionless velocity profiles for steady flows of a 60% suspension, at various rotational velocities $\Omega$. (a) Δρ = 0; the dashed line is the theoretical velocity profile for a Newtonian fluid. (b) With Δρ = 0.15 g.cm$^{-3}$.

The MRI velocity profiles reveal an important difference between suspensions that are density-matched and those that are not (Fig.1). The latter shows that we observe marked shear banding for the lowest $\Omega$, that is not present for the density matched system. Since the shear stress varies in the Couette cell as $\sigma(r) = \sigma(R_i)R_i^2/r^2$, the transition between flow and no flow directly demarcates the yield stress [16]. Importantly, the experiments also indicate the existence of a critical shear rate: the velocity profile falls down abruptly to 0 with a slope different from 0 at the interface, i.e. there is a shear rate discontinuity between the sheared and the unsheared zone. This defines the critical shear rate $\dot{\gamma}_c$ $\approx 0.05 \pm$ 0.02 s$^{-1}$ below which no stable flow exists.

Our observations point out for the first time that a slight density mismatch has most probably been at the origin of all previous observations of yield stresses and shear banding in granular suspensions. Sedimentation or creaming may lead to the creation of a dense zone in which the particles are sufficiently densely packed that a yield stress emerges. This is confirmed by our direct MRI measurements of the density profiles of the density matched and mismatched systems (Fig.2).

We observe that the density matched system is perfectly homogeneous, but for Δρ = 0.15 g.cm$^{-3}$ at rest there is significant creaming with a velocity of the order of 20 $\mu m$/s, leading to a material of 63% volume fraction. In the absence of flow there are no other interactions between noncolloidal particles than contact interactions and the existence of a yield stress can only be ascribed to the formation of a jammed contact network: the yield stress emerges around $\varphi$=63%. This directly shows that gravity plays two roles: (i) it allows for the creation of this contact network thanks to the creaming of the system; (ii) it provides the normal forces that are necessary to stabilize the granular system. The latter observation solves the problem of the observation of yield stresses without apparent normal forces [4]: the latter were in fact present due to a slight density mismatch.

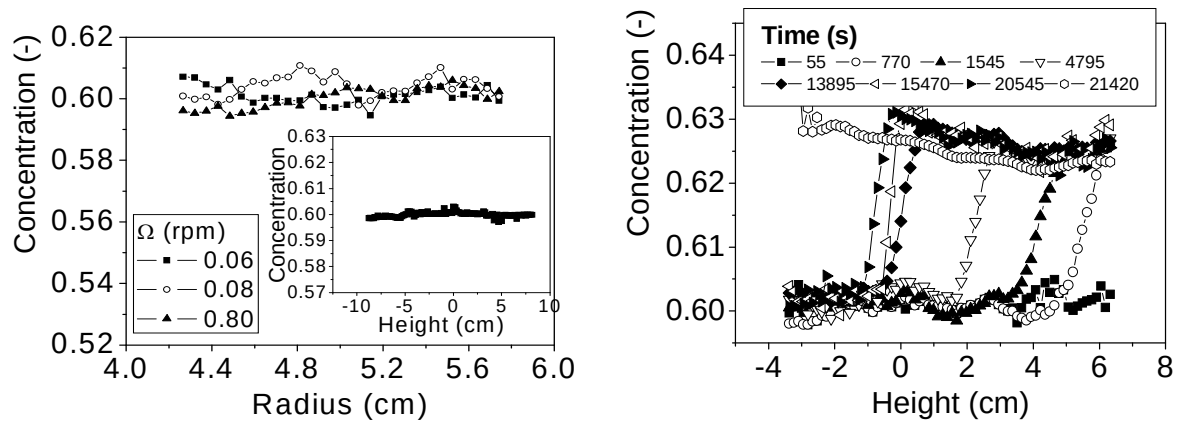

Fig. 2: (a) Radial concentration profile during flow for Δρ = 0.15 g.cm$^{-3}$. Insert: Vertical concentration profile during flow for Δρ = 0.0 g.cm$^{-3}$. (b) Time evolution of the concentration under zero shear for the Δρ = 0.15 g.cm$^{-3}$ suspension.

It turns out that when everything is flowing, the system becomes homogeneous again: there is shear induced resuspension of the particles [21] that creates normal forces that in turn lead to a particle flux opposed to that of creaming or sedimentation [6,9]. Therefore, in our system the yield stress and critical shear rate are closely related, and both find their origin in the gravitational forces. This provides a theoretical limit for the emergence of a yield stress, and also implies that shear banding appears when normal stresses generated by the flow can no longer balance gravity forces. In dense suspensions, the normal stresses are predicted to be of the same order of magnitude and to diverge like the viscous shear stresses as the volume fraction is increased [10]. The transition between the sheared and unsheared zone should then correspond to a simple balance between gravitational and viscous stresses: $\eta\dot{\gamma} = \Delta\rho g R$ where $\eta$ is the macroscopic viscosity of the suspension. Interestingly, this analysis predicts that the yield stress is accompanied by a shear-banding phenomenon *even in a homogeneous stress field*, i.e. a critical shear rate below which no flow is observed $\dot{\gamma}_c = \Delta\rho g R / \eta$, akin to what is observed for thixotropic gels [8,11]. The critical shear rate from the MRI is in very good agreement with the simple theory we provide above. With $\eta$=1Pa.s, the macroscopically measured viscosity of the paste, we find $\dot{\gamma}_c$=0.03 s$^{-1}$. Note moreover that (i) the predicted $1/\eta$ scaling of $\dot{\gamma}_c$ is in agreement with the findings of [4] who varied the interstitial liquid viscosity over 3 decades, (ii) that by varying $\Delta\rho$ we show that the $\Delta\rho$ scaling is also in agreement with the experimental results (Fig.3b). All these observations provide strong evidence that shear banding finds its origin in a competition between creaming and shear-induced resuspension. The mechanism we propose is reminiscent of the one that drives the onset of erosion of granular beds by water in rivers [13,21].

This closely resembles what is observed for thixotropic gels [22]; for these there is a competition between gel formation at rest and destruction by flow that leads to thixotropy [11]. For the granular paste, the response is also thixotropic due to the competition between creaming and resuspension. It implies the existence of a bifurcation of the viscosity: the material abruptly passes from a low viscosity flowing state above a critical (yield) stress $\sigma_c$ to a jammed state below $\sigma_c$ [11,17] and dry granular materials [28]. We thus study the macroscopic shear rate in time when a constant stress is applied. In Fig. 3(a), we indeed observe the viscosity bifurcation: below a yield stress $\sigma_c$ the viscosity diverges (the material stops flowing) while it tends towards a low finite value for stresses higher than $\sigma_c$; the shear rate is then always higher than a critical value. The critical shear rate $\dot{\gamma}_c$ = 0.13s$^{-1}$ $\pm$ 0.07 (error indicates reproducibility) measured here macroscopically is again consistent with the critical shear rate from the MRI.

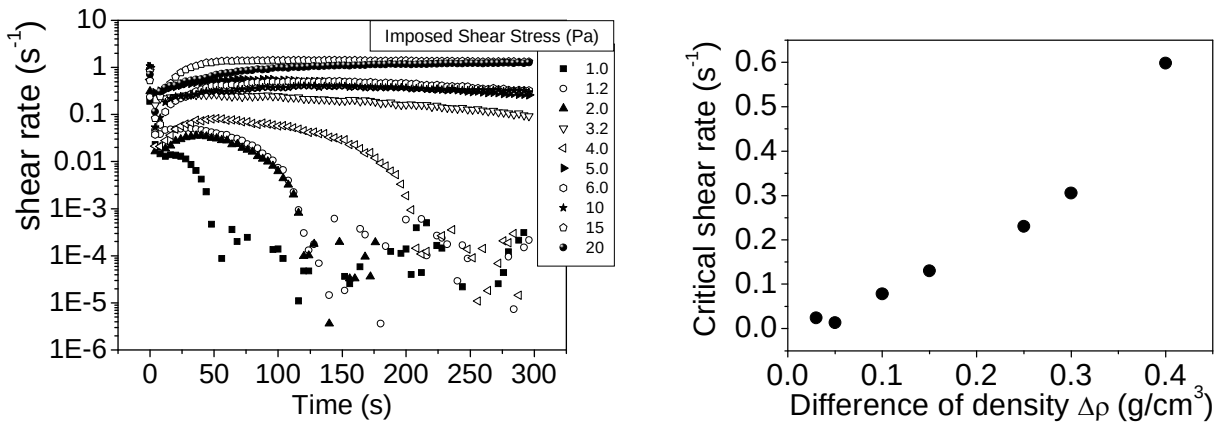

Fig. 3: a) Temporal evolution of the shear rate for different applied shear stresses in suspension with Δρ = 0.15 g.cm$^{-3}$; b) critical shear rate as a function of Δρ.

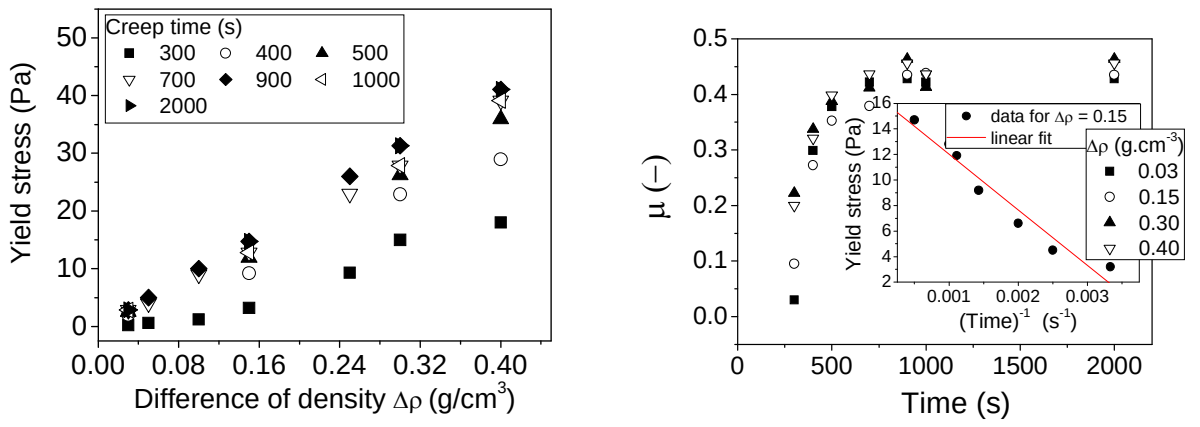

Fig. 4: (a) Evolution of the yield stress as a function the difference of density Δρ and for different times. (b) Time evolution of the macroscopic friction coefficient for different Δρ. Inset: Yield stress vs 1/time for Δρ = 0.15 g.cm$^{-3}$.

The bifurcation stress then corresponds to the effective yield stress of the system, which consequently depends both on time and on the density difference. However, true jamming would pertain to the existence of a yield stress for infinite time and no density difference. We therefore plot the time- and density dependence of the yield stress in Fig.4. It follows, as expected from our considerations above, that $\sigma_c^{t\to\infty} \propto \Delta\rho$. It shows that this yield stress can actually be ascribed completely to the frictional behavior of the granular matrix under normal stresses due to gravity: once the material has settled down, there is a contact network which has a (granular) frictional behavior characterized by a friction coefficient $\mu$, the ratio of the shear stress to the normal stress. The (frictional) yield stress thus reads $\mu\Delta\rho g z$, leading to $\sigma_c^{t\to\infty} = \frac{1}{2}\mu\Delta\rho g H$, with H the height of the cylinder of the Couette geometry. In Fig. 4(a) we indeed find a straight line with a slope corresponding to $\mu \approx 0.48$. The time evolution of the macroscopic friction coefficient μ is plotted in Fig.4(b) from which it is evident that the friction is close to zero when the particles have had no time to sediment. This indicates that there are no frictional contacts between the particles: they are not touching. This ties in with recent work on the jamming transition, where a discontinuous jump from zero to a finite number of contacts happens AT the jamming transition [23]. The surprising conclusion from Fig.4 is therefore clearly that even a very dense suspension (60%) does NOT have a true yield stress and is therefore NOT jammed. This poses the question where the jamming transition actually is. To answer this question, suspensions of different volume fraction were prepared and subjected to a very low shear rate of 0.005s$^{-1}$. The resulting stresses and strains were measured (Fig.5). For concentrations up to 62.4%, the suspension flows freely: there is no yield stress. At or slightly above 62.7%, the suspensions have a yield stress: the stress levels are very high, and initially a linear stress-strain relation is observed, as for an elastic solid.

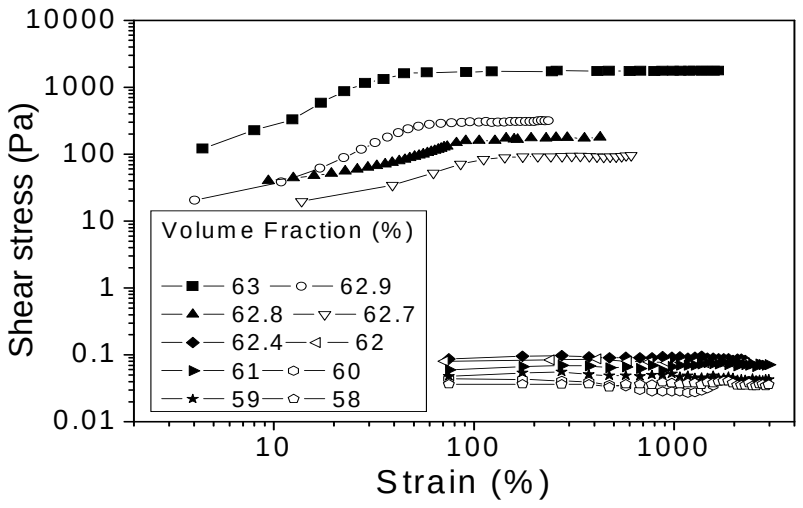

Fig. 5: Shear stress vs. shear strain for various volume fractions.

What is then the origin of the 'true' yield stress for very dense suspensions? The most likely origin for the emergence of a yield stress for Δρ = 0 is the dilatant behavior of the granular material at high volume fractions. When the density of the granular material is high enough, it needs to dilate (Reynolds' dilatancy) in order to flow; however this is not always possible due to confinement by the fluid [7]. The effect of confinement can be evaluated: the typical normal force exerted on the granular skeleton is due to surface tension and is thus here of order $\gamma/R \approx 1000$ Pa, with $\gamma \approx 20 mNm^{-1}$ the surface tension of the suspending liquid. This would give a (frictional) yield stress of order 100 to 1000 Pa (depending on $\mu$) in good agreement with the measured shear stresses of Fig. 5; in fact the plateau value of the stress at high strains is often taken as a good measure of the yield stress [17]. This simple picture predicts that the jamming transition in suspensions occurs at the critical state density. For almost frictionless ($\mu$<0.01) dry granular materials, this density was found to be 62.5 $\pm$ 0.5% [24], in excellent agreement with the value found here. However, theoretically this density was recently found to be equal to 63.9% for frictionless grains [14], and decreases very quickly when the intergrain friction is increased [24]. From simulations [24], a 62.7% volume fraction would correspond to a very low intergrain friction coefficient $\mu$=0.05, pointing out the important lubricating role of the interstitial fluid.

In conclusion, we have shown that the competition between gravity and viscous stresses is at the origin of the development of a dynamic yield stress and of shear banding in granular suspensions at relatively low volume fractions. However, when the gravity forces play no role, we have shown that no yield stress is encountered until a volume fraction of 62.7 $\pm$ 0.3%, that is significantly lower than the random close packing. The simple mechanism at play here for shear banding may be seen more generally as a competition between structuration and destructuration, a phenomenon which seems to be a hallmark of systems that exhibit shear banding [8,16-18]; our system may then be a good simple model system for tests of shear banding pictures.

Interestingly, our results may also be connected to recent observations in hard-sphere colloidal glasses. It is generally assumed that the glass transition in these systems happens at a volume fraction of 58% [24]. However recent experiments in space were not able to see a glass transition at all around these volume fractions [25], very similarly to what we see here for a system that is only different from the hard-sphere colloids in the sense that there is no Brownian motion. In addition, for hard-sphere colloids, recent experiments using confocal microscopy [25] report an accelerated aging under gravity that is very much compatible with the creaming or sedimentation observed here. It may thus very well be that the true 'jamming' (glass transition) point of colloidal hard spheres is close to the jamming point of non colloidal hard spheres reported here.